\documentclass[12pt, twoside]{article}

\usepackage{a4wide, amsmath, latexsym, epsfig,
  psfrag,graphicx, rotating, fancyhdr, tabularx, afterpage}

\usepackage{feynarts}
\usepackage{axodraw}

\newcommand{\id}{{\rm 1\kern-.12em
\rule{0.3pt}{1.5ex}\raisebox{0.0ex}{\rule{0.1em}{0.3pt}}}}

\newcommand{\nl}{\nonumber\\}

\newcommand{\lpar}{\left(}                            
\newcommand{\rpar}{\right)}

\newcommand{\bq}{\begin{equation}}                    
\newcommand{\eq}{\end{equation}}
\newcommand{\bqa}{\arraycolsep 0.14em\begin{eqnarray}}
\newcommand{\eqa}{\end{eqnarray}}
\newcommand{\ba}[1]{\begin{array}{#1}}
\newcommand{\ea}{\end{array}}
\newcommand{\ben}{\begin{enumerate}}
\newcommand{\een}{\end{enumerate}}
\newcommand{\bei}{\begin{itemize}}
\newcommand{\eei}{\end{itemize}}

\newcommand{\fig}[1]{Fig.~\ref{#1}}

\def\Re{\mathop{\operator@font Re}\nolimits}
\def\Im{\mathop{\operator@font Im}\nolimits}

\newcommand{\spro}[2]{{#1}\cdot{#2}}
\newcommand{\egam}[1]{\Gamma\lpar#1\rpar}               

\newcommand{\upar}[1]{u}

\newcommand{\bqas}{\begin{eqnarray*}}
\newcommand{\eqas}{\end{eqnarray*}}

\def\app#1#2 {{\it Acta. Phys. Pol.} {\bf#1},#2}
\def\cpc#1#2 {{\it Computer Phys. Comm.} {\bf#1},#2}
\def\np#1#2 {{\it Nucl. Phys.} {\bf#1},#2}
\def\pl#1#2 {{\it Phys. Lett.} {\bf#1},#2}
\def\prep#1#2 {{\it Phys. Rep.} {\bf#1},#2}
\def\prev#1#2 {{\it Phys. Rev.} {\bf#1},#2}
\def\prl#1#2 {{\it Phys. Rev. Lett.} {\bf#1},#2}
\def\zp#1#2 {{\it Zeit. Phys.} {\bf#1},#2}
\def\sptp#1#2 {{\it Suppl. Prog. Theor. Phys.} {\bf#1},#2}
\def\mpl#1#2 {{\it Modern Phys. Lett.} {\bf#1},#2}
\def\jetp#1#2 {{\it Sov. Phys. JETP} {\bf#1},#2}
\def\fpj#1#2 {{\it Fortschr. Phys.} {\bf#1},#2}
\def\afp#1#2 {{\it Acta.Phys. Polon.} {\bf#1},#2}
\def\err#1#2 {{\it Erratum} {\bf#1},#2}
\def\ijmp#1#2 {{\it Int. J. Mod. Phys} {\bf#1},#2}
\def\nc#1#2 {{\it Nuovo Cimento} {\bf#1},#2}
\def\ap#1#2 {{\it Ann. Phys.} {\bf#1},#2}
\def\cmp#1#2 {{\it Comm. Math. Phys.} {\bf#1},#2}
\def\el#1#2 {{\it Europhys. Lett.} {\bf#1},#2}
\def\hpa#1#2 {{\it Helv. Phys. Acta} {\bf#1},#2}
\def\yf#1#2 {{\it Yad. Fiz.} {\bf#1},#2}
\def\nim#1#2 {{\it Nucl. Instrum. Meth.} {\bf#1},#2}
\def\spz#1#2 {{\it Sov. Pisma Zhetf} {\bf#1},#2}
\def\jetpl#1#2 {{\it JETP Lett.} {\bf#1},#2}
\def\sjnp#1#2 {{\it Sov. J. Nucl. Phys.} {\bf#1},#2}
\def\ptp#1#2 {{\it Progr. Theor. Phys. (Kyoto)} {\bf#1},#2}
\def\rmp#1#2  {{\it Rev. Mod. Phys.} {\bf#1},#2}
\def\zhetf#1#2 {{\it ZhETF} {\bf#1},#2}
\def\prs#1#2 {{\it Proc. Roy. Soc.} {\bf#1},#2}
\def\phys#1#2 {{\it Physica} {\bf#1},#2}

\def\bfi{\begin{figure}}
\def\efi{\end{figure}}

\newcommand{\intsx}[1]{\int_{\scriptstyle 0}^{\scriptstyle 1}\!\!\!d#1}
\newcommand{\intsxy}[2]{\int_{\scriptstyle 0}^{\scriptstyle 1}\!\!\!d#1
                        \int_{\scriptstyle 0}^{\scriptstyle #1}\!\!\!d#2}

\newcommand{\seff}{\sin^2\theta_{\rm eff}}

\begin{document}

\thispagestyle{empty}
\setcounter{page}{0}
\def\thefootnote{\fnsymbol{footnote}}

{\textwidth 15cm

\begin{flushright}
MPP-2005-77\\
hep-ph/0507158 \\
\end{flushright}

\vspace{2cm}

\begin{center}

{\Large\sc {\bf The effective electroweak mixing angle 
                \boldmath{$\sin^2\theta_{\rm eff}$} \\[0.3cm]
               with two-loop fermionic contributions}}

\vspace{2cm}

{\sc W. Hollik, U. Meier} and {\sc S. Uccirati}

\vspace*{1cm}

     Max-Planck-Institut f\"ur Physik \\[0.1cm] 
     (Werner-Heisenberg-Institut)\\[0.1cm]
     D-80805 M\"unchen, Germany
\end{center}

\vspace*{2cm}

\begin{abstract}
\noindent
We present the results from a calculation of the 
full electroweak two-loop fermionic contributions
to the effective leptonic mixing angle of the $Z$ boson,
$\sin^2\theta_{\rm eff}$, in the Standard Model.
On-shell renormalization and analytic calculations are
performed for the three-point vertex functions at
zero external momenta,
whereas 
irreducible three-point integrals for non-vanishing external
momenta are evaluated semi-analytically applying 
two different methods.
Comparisons with a previous calculation 
show complete agreement. 
\end{abstract}

}
\def\thefootnote{\arabic{footnote}}
\setcounter{footnote}{0}

\newpage

\section{Introduction}
Precision observables at the $Z$-boson resonance, measured at LEP 1 and SLC, 
together with the masses of the $W$ boson  and of the top quark, measured at
LEP 2 and the Tevatron, constitute a set of high-energy quantites which in
comparison with the Standard Model predictions allow global precision analyses
yielding indirect bounds on the Higgs-boson mass~$M_H$~\cite{unknown:2004qh}. 
In this context, one of the most important quantities with a very high sensitivity
to $M_H$ is the effective leptonic mixing angle, expressed as
$\seff$, in the effective couplings of the $Z$ to leptons (electrons, to be precise).
Besides $M_W$, it is a key observable measured with high experimental accuracy, 
with expected further improvements in the future.
The current experimental value is 
$0.23147 \pm 0.00017$~\cite{unknown:2004qh}; a linear electron--positron collider 
with GigaZ capabilities could even reach an accuracy of 
$1.3 \times 10^{-5}$~\cite{Aguilar-Saavedra:2001rg,Baur:2001yp}. 
Therefore, very precise and reliable 
theoretical predictions for $\seff$ are required.

The electroweak mixing angle in the effective leptonic vertex of the $Z$ boson
can be defined via the relation (see e.g.~\cite{Bardin:1997xq})
\begin{eqnarray}
\seff &=& \frac{1}{4}  \left( 1-\mathrm{Re}\, \frac{g_V}{g_A} \right) , \label{s2w}
\end{eqnarray}
in terms of the dressed vector and axial vector couplings 
$g_{V,A}$ in the $Z$--lepton vertex.
Starting from the $Z$ mass $M_Z$ and the $W$ mass $M_W$, 
$\seff$ can be obtained in the  following way,
\begin{eqnarray}
\seff &=& \kappa \, s_W^2 \; = \; 
\kappa \,\left( 1-\frac{M_W^2}{M_Z^2} \right) \, , \quad  
 \kappa = 1+\Delta\kappa \, , \label{kappa}
\end{eqnarray}
with higher-order contributions collected in the quantity $\Delta\kappa$.
The $W$ mass can be related to the precisely known Fermi constant $G_F$ with 
the help of the $W$--$Z$ interdependence,
\begin{eqnarray}
 M_W^2 \left( 1-\frac{M_W^2}{M_Z^2} \right) &=& 
       \frac{\pi\alpha}{\sqrt{2} G_F} (1+\Delta r) \, , 
\end{eqnarray}
where $\Delta r$ denotes the higher-order contributions. Recently, the
full electroweak two-loop calculation for $\Delta r$
has been completed, with both fermionic
loops~\cite{Freitas:2002ja,Awramik:2003ee} 
and purely bosonic loops~\cite{Awramik:2002wn}, 
and further improvement by  three-loop
contributions to the $\rho$-parameter~\cite{Faisst:2003px}
and to the $S$ parameter~\cite{Boughezal:2005eb}
has been given.
Moreover, the universal QCD corrections via the 
self energies~\cite{Djouadi:1987gn}  are known.
 
For the $\kappa$ factor in (\ref{kappa}), fewer higher-order terms have been 
evaluated as yet. Besides the universal contributions through the 
$\rho,S$ parameters
and the QCD corrections,
a top mass expansion for the electroweak two-loop corrections
was performed~\cite{Degrassi:1996ps}, and only
recently the complete two-loop electroweak corrections of the fermionic type, i.e.\
with at least one closed 
fermion loop, were calculated~\cite{Awramik:2004ge}. In this paper we present 
an independent calculation of those fermionic two-loop contributions and 
perform a  comparison with the result of~\cite{Awramik:2004ge}.

\section{Electroweak two-loop contributions}\label{2loop}
\begin{figure}[!b]
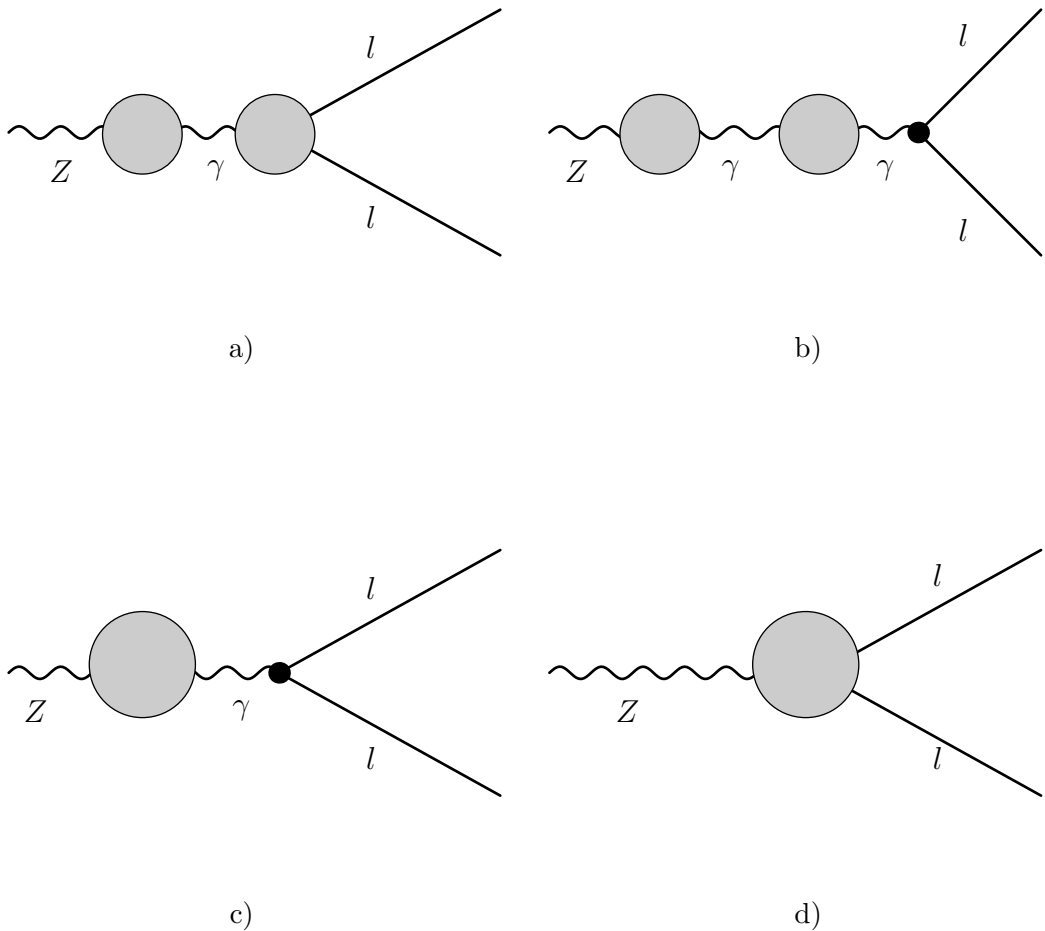


\begin{feynartspicture}(450,450)(2,2.2)
\FADiagram{a)}
\FAProp(0.,10.)(11.,10.)(0.,){/Sine}{0}
\FALabel(9.,7.43)[t]{$\gamma$}
\FALabel(3.,7.43)[t]{$Z$}
\FAProp(20.,15.)(11.,10.)(0.,){/Straight}{0}
\FALabel(15.2273,11.3749)[br]{$l$}
\FAProp(20.,5.)(11.,10.)(0.,){/Straight}{0}
\FALabel(15.2273,5.62506)[tr]{$l$}
\FAVert(11.,10.){0}
\GCirc(130.,305.){15.}{0.8}
\GCirc(80.,305.){15.}{0.8}

\FADiagram{b)}
\FAProp(0.,10.)(15.,10.)(0.,){/Sine}{0}
\FALabel(13.,7.43)[t]{$\gamma$}
\FALabel(7.,7.43)[t]{$\gamma$}
\FALabel(1.,7.43)[t]{$Z$}
\FAProp(20.,15.)(15.,10.)(0.,){/Straight}{0}
\FALabel(16.2273,11.8749)[br]{$l$}
\FAProp(20.,5.)(15.,10.)(0.,){/Straight}{0}
\FALabel(16.2273,5.12506)[tr]{$l$}
\FAVert(15.,10.){0}
\GCirc(335.,305.){15.}{0.8}
\GCirc(275.,305.){15.}{0.8}

\FADiagram{c)}
\FAProp(0.,10.)(11.,10.)(0.,){/Sine}{0}
\FALabel(10.,8.43)[t]{$\gamma$}
\FALabel(2.,8.43)[t]{$Z$}
\FAProp(20.,15.)(11.,10.)(0.,){/Straight}{0}
\FALabel(15.2273,12.3749)[br]{$l$}
\FAProp(20.,5.)(11.,10.)(0.,){/Straight}{0}
\FALabel(15.2273,6.62506)[tr]{$l$}
\FAVert(11.,10.){0}
\GCirc(80.,105.){20.}{0.8}

\FADiagram{d)}
\FAProp(0.,10.)(11.,10.)(0.,){/Sine}{0}
\FALabel(3.,8.43)[t]{$Z$}
\FAProp(20.,15.)(11.,10.)(0.,){/Straight}{0}
\FALabel(15.2273,12.8749)[br]{$l$}
\FAProp(20.,5.)(11.,10.)(0.,){/Straight}{0}
\FALabel(15.2273,6.62506)[tr]{$l$}
\FAVert(11.,10.){0}
\GCirc(330.,105.){20.}{0.8}

\end{feynartspicture}

\hspace*{-10cm}
\caption{Classes of two loop diagrams}
\label{gendiag}
\end{figure}

Expanding the dressed couplings in (\ref{s2w}) according to
$g_{V,A} = g_{V,A}^{(0)} \left(1+g_{V,A}^{(1)}+g_{V,A}^{(2)}+\cdots\right)$ 
in powers of $\alpha$, 
one obtains the $O\left(\alpha^2\right)$ 
contribution to $\seff$ in the loop expansion 
\begin{eqnarray}
\seff &=&
\seff^{(0)}+\seff^{(1)}+
\seff^{(2)}+\mathcal{O}\left(\alpha^3\right)
\end{eqnarray}
to be
\begin{eqnarray}
\seff^{(2)} &=& -\frac{g_V^{(0)}}{4 g_A^{(0)}} \;
\mathrm{Re}\left(g_V^{(2)}-g_A^{(2)}+g_A^{(1)} 
\left( g_A^{(1)}-g_V^{(1)} \right)\right).\label{expand}
\end{eqnarray}
Hence, besides the two-loop diagrams depicted schematically in Fig.~\ref{gendiag}, 
products of one-loop contributions have to be taken into account. 
The circles in the diagrams of  Fig.~\ref{gendiag} denote the renormalized
two- and three-point vertex functions, 
at the one-loop level in the reducible
diagrams of Fig.~\ref{gendiag}a and \ref{gendiag}b, and at two-loop order   
in the irreducible diagrams of Fig.~\ref{gendiag}c and \ref{gendiag}d.

In our computation we have used the on-shell renormalization scheme as described 
in~\cite{Freitas:2002ja}. 
In this scheme the real part of the diagram shown in Fig.~\ref{gendiag}c 
vanishes. 
The reducible diagrams of Fig.~\ref{gendiag}a and \ref{gendiag}b
only contribute products of imaginary parts of one-loop functions, 
which can be easily computed. So the genuine two-loop task is the
calculation of the irreducible $Z\ell\ell$-vertex diagrams 
in Fig.~\ref{gendiag}d.

\begin{figure}[!htb]
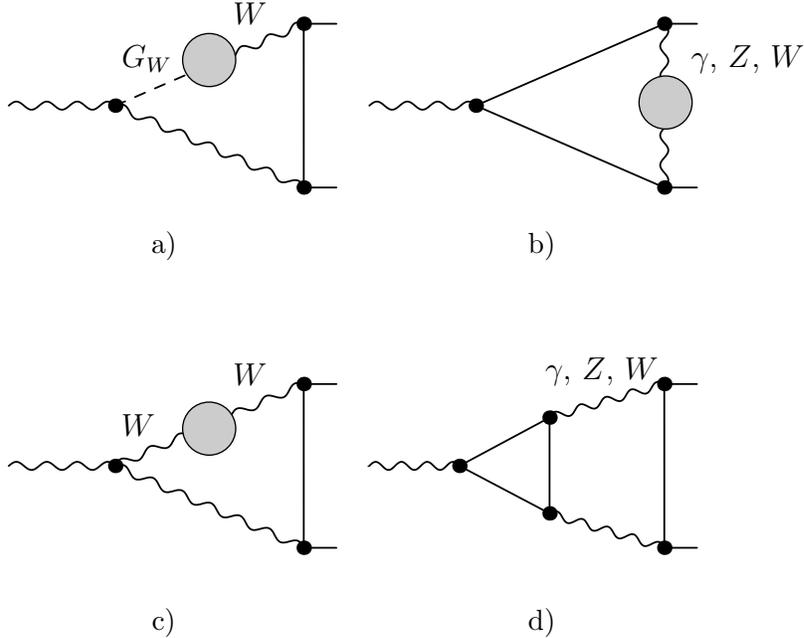

\begin{feynartspicture}(450,300)(2,2.2)
\FADiagram{a)}
\FAProp(0.,10.)(6.5,10.)(0.,){/Sine}{0}
\FAProp(20.,15.)(18.,15.)(0.,){/Straight}{0}
\FAProp(20.,5.)(18.,5.)(0.,){/Straight}{0}

\FAProp(12.25,12.5)(18.,15.)(0.,){/Sine}{0}
\FAProp(6.5,10.)(12.25,12.5)(0.,){/ScalarDash}{0}

\FAProp(6.5,10.)(18.,5.)(0.,){/Sine}{0}
\FAProp(18.,15.)(18.,5.)(0.,){/Straight}{0}
\FAVert(6.5,10.){0}
\FAVert(18.,15.){0}
\FAVert(18.,5.){0}
\FALabel(9., 12.)[t]{$G_W$}
\FALabel(15., 14.5)[t]{$W$}
\GCirc(170.,221.){10.}{0.8}

\FADiagram{b)}

\FAProp(0.,10.)(6.5,10.)(0.,){/Sine}{0}
\FAProp(20.,15.)(18.,15.)(0.,){/Straight}{0}
\FAProp(20.,5.)(18.,5.)(0.,){/Straight}{0}

\FAProp(6.5,10.)(18.,15.)(0.,){/Straight}{0}
\FAProp(6.5,10.)(18.,5.)(0.,){/Straight}{0}
\FAProp(18.,15.)(18.,5.)(0.,){/Sine}{0}
\FAVert(6.5,10.){0}
\FAVert(18.,15.){0}
\FAVert(18.,5.){0}
\FALabel(22., 12.)[t]{$\gamma$, $Z$, $W$}
\GCirc(342.,205.){10.}{0.8}

\FADiagram{c)}

\FAProp(0.,10.)(6.5,10.)(0.,){/Sine}{0}
\FAProp(20.,15.)(18.,15.)(0.,){/Straight}{0}
\FAProp(20.,5.)(18.,5.)(0.,){/Straight}{0}

\FAProp(6.5,10.)(18.,15.)(0.,){/Sine}{0}

\FAProp(6.5,10.)(18.,5.)(0.,){/Sine}{0}
\FAProp(18.,15.)(18.,5.)(0.,){/Straight}{0}
\FAVert(6.5,10.){0}
\FAVert(18.,15.){0}
\FAVert(18.,5.){0}
\FALabel(8.5, 12.5)[t]{$W$}
\FALabel(15., 15.5)[t]{$W$}
\GCirc(170.,82.){10.}{0.8}

\FADiagram{d)}

\FAProp(0.,10.)(5.5,10.)(0.,){/Sine}{0}
\FAProp(20.,15.)(18.,15.)(0.,){/Straight}{0}
\FAProp(20.,5.)(18.,5.)(0.,){/Straight}{0}
\FAProp(11.,7.1)(5.5,10.)(0.,){/Straight}{0}
\FAProp(11.,7.1)(18.,5.)(0.,){/Sine}{0}
\FAProp(11.,12.95)(11.,7.1)(0.,){/Straight}{0}
\FAProp(11.,12.95)(5.5,10.)(0.,){/Straight}{0}
\FAProp(11.,12.95)(18.,15.)(0.,){/Sine}{0}
\FAProp(18.,15.)(18.,5.)(0.,){/Straight}{0}
\FAVert(11.,12.95){0}
\FAVert(11.,7.1){0}
\FAVert(5.5,10.){0}
\FAVert(18.,15.){0}
\FAVert(18.,5.){0}
\FALabel(13.5,15.8)[t]{$\gamma$, $Z$, $W$}

\end{feynartspicture}
\caption{Irreducible two-loop vertex diagrams containing fermion loops.}
\label{diag}
\end{figure}

For the computation of
the two-loop corrections with at least one closed fermion loop 
we are directed to the
vertex diagrams shown in Fig.~\ref{diag}, 
one-loop diagrams with one-loop counter term insertions of the fermionic type, 
and the two-loop fermionic counter term.
In the combination entering (\ref{expand}),
the two-loop counter term  for $\seff$
in the on-shell scheme is given by 

\begin{eqnarray}
\label{deltasineff} 
\delta \seff^{(2)}   &=& \delta s^2_W+ 
\frac{1}{2 s_W c_W}\delta Z^{\gamma Z}_{(2)}+s^2_W \left(s^2_W-c^2_W\right) \delta Z^{\ell,L}_{(2)} -s^2_W \left(s^2_W-c^2_W\right) \delta Z^{\ell,R}_{(2)}\label{2lCT}\\[0.2cm]
\nonumber && \quad \quad \quad \quad \quad \quad 
 +\;\mathrm{products}\;\mathrm{of}\; 
\rm{one}\; \rm{loop} \;\mathrm{counter}\;\mathrm{terms}.
\end{eqnarray}

$\delta Z^{\gamma Z}_{(2)}$ and $\delta Z^{\ell,L/R}_{(2)}$ are the field renormalization 
constants for the photon-$Z$ two-point function and for 
the left-/right-handed lepton field.
The counter term 
$\delta s^2_W$ 
for the on-shell quantity 
$s^2_W$  
in (\ref{kappa})
contains the two-loop counter terms for
the $Z$- and $W$-boson masses, $\delta M^2_{Z/W(2)}$,
and products of one-loop counter terms,
\begin{eqnarray}
\delta s^2_W &=& 
\frac{M^2_W}{M^2_Z}\left(\frac{\delta M^2_{Z(2)}}{M^2_Z}
-\frac{\delta M^2_{W(2)}}{M^2_W}\right)
- \frac{M^2_W}{M^2_Z}\frac{\delta M^2_{Z(1)}}{M^2_Z}
\left(\frac{\delta M^2_{Z(1)}}{M^2_Z}-\frac{\delta M^2_{W(1)}}{M^2_W}\right).
\label{ds2w2}
\end{eqnarray}

\newpage

Apart from the products of one-loop functions in (\ref{expand}) and the reducible diagram 
Fig.~\ref{gendiag}b only the two-loop counter term (\ref{2lCT}) yields contributions 
with two closed fermion loops.

The contributions with one closed fermion loop contain the generic two-loop diagrams 
depicted in Fig.~\ref{diag}.
Their evaluation encounters a twofold task:
the two-loop renormalization, and the computation of genuine two-loop vertex functions 
for non-vanishing external momenta.
We have separated these two items
by splitting the renormalized
two-loop contribution to the $Z\ell\ell$ vertex function
with the two external lepton momenta $p_{\pm}$,
\begin{eqnarray}
\hat{\Gamma}_\mu^{Z\ell\ell\,(2)}(P^2) =
\gamma_\mu \left( g_V^{(2)} -g_A^{(2)} \gamma_5\right) , \quad \quad
P^2 \;=\; (p_- + p_+)^2\, ,
\end{eqnarray}
into two UV-finite pieces according to 
\begin{eqnarray}
\hat{\Gamma}_\mu^{Z\ell\ell\,(2)}(M^2_Z) &=& 
\Gamma_\mu^{Z\ell\ell\,(2)} (M^2_Z) + \Gamma_\mu^{CT}
\nonumber \\[0.2cm]
 &=&  
\left[ \Gamma_\mu^{Z\ell\ell\, (2)}(0)+ \Gamma_\mu^{CT}\right]
+\left[ \Gamma_\mu^{Z\ell\ell\,(2)}(M^2_Z) - 
\Gamma_\mu^{Z\ell\ell\,(2)}(0) \right] .\label{split}
\end{eqnarray}
$\Gamma_\mu^{Z\ell\ell\,(2)}\left(P^2\right)$ 
denotes the corresponding unrenormalized  $Z\ell\ell$ vertex, 
and $\Gamma_\mu^{CT}$ is the two-loop counter term, which is independent of $P^2$.
The first term on the right hand side of (\ref{split}) contains the complete two-loop 
renormalization, but no genuine two-loop vertex diagrams since in absence of external 
momenta they reduce to simpler vacuum integrals. 
All the genuine two-loop vertex diagrams appear as subtracted quantities in
the second term on the right hand side of (\ref{split}).


\section{Outline of the calculation}\label{outline}

In order to calculate the irreducible vertex diagrams 
of Fig.~\ref{diag}, various techniques are applied.
Diagrams containing 
a one-loop self-energy insertion are calculated with the help of dispersion relations.
This results in one-dimensional integrals, which are computed 
numerically.

The diagrams with fermion triangles, Fig.~\ref{diag}d, 
require a careful treatment of $\gamma_5$  
in $D$ dimensions. 
In \cite{Freitas:2002ja} it was shown that naive dimensional regularization 
with a four-dimensional treatment of the $\epsilon$-tensor from traces 
involving $\gamma_5$ 
can be applied for this special case.
Putting such traces to zero yields a UV-finite difference, which
vanishes for a fermion generation with equal masses.
Since the light fermion masses have been neglected in our approach,
a non-vanishing contribution occurs only from the third generation.
In the numerical discussion of section \ref{results} this difference
is labeled as the "$\gamma_5$" contribution.

Differently from the charged-current case treated in \cite{Freitas:2002ja},
the neutral-current vertex considered here involves another subtlety
originating from photons in the internal lines of Fig.~\ref{diag}d,
which give rise to collinear divergences in the case of massless electrons.
This would produce unphysical finite extra shifts in combination with the
$(D-4)$ terms from the $\gamma_5$ traces. [For massive internal bosons
those terms disappear in the limit $D\rightarrow 4$ because the second
loop integration does not introduce another UV divergence.]  
In order to prevent this kind of problems, we keep
the physical mass of the external electrons in the dangerous diagrams.
Moreover, also  the masses of the internal $b$ quarks 
and $\tau$ leptons have been kept.
The resulting two-loop vertex functions are further evaluated
using the methods described 
in~\cite{Ferroglia:2003yj}. In the end we are left with up to four-dimensional integrals,
which are calculated numerically.
%

\subsection{Computation of \boldmath{$\hat{\Gamma}_\mu^{Z\ell\ell\,(2)}(0)$} } \label{ct}

In order to get the renormalized vertex (respectively,  the contribution to $\seff$)
from the two-loop vertex at zero momentum, the two-loop renormalization
constants in (\ref{deltasineff}) and (\ref{ds2w2}) are needed. 
In the on-shell scheme~\cite{Freitas:2002ja} they are given by
\begin{eqnarray}
\delta M^2_{W(2)} &=& \mathrm{Re} \left(\Sigma^{W}_{(2)}(M^2_W)\right)
-\delta Z^{W}_{(1)}\delta M^2_{W(1)}+\mathrm{Im}\left(\Sigma^{W'}_{(1)}(M^2_W)\right)
\mathrm{Im}\left(\Sigma^{W}_{(1)}(M^2_W)\right) , \label{dmw2}\\
\nonumber\delta M^2_{Z(2)} &=& \mathrm{Re}\left(\Sigma^{ZZ}_{(2)}(M^2_Z)\right)
-\delta Z^{ZZ}_{(1)}\delta M^2_{Z(1)}+\mathrm{Im}\left(\Sigma^{ZZ'}_{(1)}(M^2_Z)\right)
\mathrm{Im}\left(\Sigma^{ZZ}_{(1)}(M^2_Z)\right)\\
&&+\frac{M^2_Z}{4}\left(\delta Z^{\gamma Z}_{(1)}\right)^2+\frac{\mathrm{Im}
\left(\Sigma^{\gamma Z}_{(1)}(M^2_Z)\right)^2}{M^2_Z},\\
\delta Z^{\gamma Z}_{(2)} &=& -2 \frac{\mathrm{Re}
\left(\Sigma^{\gamma Z}_{(2)}(M^2_Z)\right)}{M^2_Z}
+\delta Z^{Z\gamma}_{(1)}\, \frac{\delta M^2_{Z(1)}}{M^2_Z}
-\frac{1}{2}\delta Z^{\gamma Z}_{(1)}\delta Z^{\gamma\gamma}_{(1)},\label{dZA2}
\end{eqnarray}
involving the bosonic self energies and field renormalization constants, and 
\begin{eqnarray}
\delta Z^{\ell,L}_{(2)} &=& -\Sigma^{\ell,L}_{(2)}(0),\label{dzl2}\\
\delta Z^{\ell,R}_{(2)} &=& -\Sigma^{\ell,R}_{(2)}(0) \, ,\label{dzr2}
\end{eqnarray}
from the two-loop leptonic self energies according to the decomposition
\begin{eqnarray}
\Sigma^\ell(p) = \not{\!p}\, \omega_L \Sigma^{\ell L}(p^2) +
                 \not{\!p} \, \omega_R \Sigma^{\ell R}(p^2)  \, .
\end{eqnarray}
$\Sigma' \left(p^2\right)$ denotes the derivative of a self-energy $\Sigma\left(p^2\right)$ 
with respect to $p^2$.

The computation of the renormalized vertex at vanishing external momentum is  done in the 
following steps.
The Feynman diagrams are generated with the package {\it FeynArts}~\cite{Hahn:2000jm}. 
The program {\it TwoCalc}~\cite{Weiglein:1993hd} is applied to reduce the amplitudes to standard 
integrals. The resulting scalar one-loop integrals and two-loop vacuum integrals
are calculated using  
analytic results \cite{'tHooft:1978xw,Davydychev:1992mt}. The two-loop self-energy 
functions with non-vanishing external momenta appearing in  the renormalization 
constants (\ref{dmw2})-(\ref{dZA2}) are obtained with the help of
one-dimensional integral representations \cite{Bauberger:1994nk}.

\subsection{Diagrams with self-energy insertions}\label{SE}
The strategy to compute vertex diagrams with self-energy insertions
for the  subtracted vertex 
$\Gamma_\mu^{Z\ell\ell}(M^2_Z) - \Gamma_\mu^{Z\ell\ell}(0)$ 
will be illustrated in terms of Fig.~\ref{diag}b with a diagonal $Z$ self-energy
as an example (the other diagrams are calculated in an analogous way).
The corresponding vertex diagram has the structure 
\begin{equation}
 \Gamma_\mu(P^2) \sim
\int {\rm d}^D \!q \; \gamma^\rho (g_v-g_a\gamma_5)
\frac{1}{\not{\!q}\,-\not{\!p_-}}\, \gamma_\mu(g_v-g_a\gamma_5) \,
\frac{1}{\not{\!q}\,+\not{\!p_+}}\, \gamma^\sigma \;
(g_v-g_a\gamma_5)
\frac{\hat{\Sigma}_{\rho\sigma}(q)}{(q^2 - M_Z^2)^2} \; ,
\label{1loop}
\end{equation}
with $P^2 =(p_- + p_+)^2$.
The renormalized  one-loop self-energy can be decomposed into
its transverse and 
longitudinal part,
\begin{eqnarray}
\hat{\Sigma}_{\rho\sigma}(q)&=& \hat{\Sigma}_T(q^2) 
\left(g_{\rho\sigma}-\frac{q_\rho q_\sigma}{q^2}\right) + \hat{\Sigma}_L(q^2) 
\left(\frac{q_\rho q_\sigma}{q^2}\right) .
\end{eqnarray}
It can be easily verified that for on-shell leptons 
the part proportional to $q_\rho q_\sigma$ yields a term independent
 of $P^2$,  hence it drops out in the difference
$\Gamma_\mu^{Z\ell\ell}(M^2_Z)- \Gamma_\mu^{Z\ell\ell}(0)$,
and only the transverse part has to be taken into account.
In the on-shell-scheme it reads
\begin{eqnarray}
\hat{\Sigma}_T(q^2) &=& \Sigma_T(q^2)
-\mathrm{Re}\, \Sigma_T(M_Z)
-(q^2-M^2_Z)\;\mathrm{Re}\, \Sigma'_T(M_Z^2) \\[0.2cm]
\nonumber &=& \Sigma_T(q^2)-\Sigma_T(M_Z^2)
-(q^2-M^2_Z)\;\Sigma'_T(M_Z^2) \\[0.1cm]
\nonumber &&+i\;\mathrm{Im}\, \Sigma_T(M_Z^2) + i\;(q^2-M^2_Z)\;
\mathrm{Im}\, \Sigma'_T(M_Z^2) \, .
\end{eqnarray}
Inserting the imaginary parts into (\ref{1loop}), one gets a constant times one-loop 
functions. For the rest, we use the dispersion relation
\begin{eqnarray}
\Sigma_T(q^2) &=&
\frac{1}{\pi}\int^{\infty}_{0} {\rm d}s\,  
\frac{\mathrm{Im}\, \Sigma_T(s)}{s-q^2 - i \epsilon} \, ,
\end{eqnarray}
yielding
\begin{eqnarray}
 & &  \Sigma_T(q^2)-\Sigma_T(M_Z^2)
-(q^2-M^2_Z)\;\Sigma'_T(M_Z^2)  \nonumber \\[0.2cm] 
 & &  = \;   
-\frac{\left(q^2-M^2_Z\right)^2}{\pi}
\int^{\infty}_{0} {\rm d}s\,
\frac{\mathrm{Im}\, \Sigma_T(s)}{(s-M_Z^2-i\epsilon)^2} \,\frac{1}{(q^2-s+i\epsilon)} \, .
\end{eqnarray}
Inserting this into (\ref{1loop}), the factor $\left(q^2-M^2_Z\right)^2$ cancels the corresponding 
term in the denominator, and  
we end up with a one-dimensional integral over a one-loop 
three-point function 
with one variable mass $s$,  multiplied by the weight-factor 
$\mathrm{Im}\Sigma\left(s\right)/(s-M_Z^2)^2$. 
The subtraction at $P^2=0$ ensures that the expressions are UV finite. 
Finally we are left with integrals of the type
\begin{eqnarray}
\int^{\infty}_{0} {\rm d}s\,
\frac{f\left(s\right)}{\left(s-M^2-i \epsilon \right)^2}  
&=& \int^{2M^2}_{0} {\rm d}s\; 
\frac{f\left(s\right)-f\left(M^2\right)-(s-M^2)f'\left(M^2\right)}{\left(s-M^2-i \epsilon \right)^2} 
\label{Int}\\ 
\nonumber 
& & + \int^{2M^2}_{0} {\rm d}s\; 
\frac{f\left(M^2\right)+(s-M^2)f'\left(M^2\right)}{\left(s-M^2-i \epsilon \right)^2} 
\\
& & + \int^{\infty}_{2M^2} {\rm d}s\;  
\frac{f\left(s\right)}{\left(s-M^2\right)^2} \, .  \nonumber
\end{eqnarray}
The second term in (\ref{Int}) can be computed analytically. 
The other terms in (\ref{Int}) are free of 
singularities and can be integrated numerically.  
We have used the CUBA library \cite{Hahn:2004fe} which allows a very fast and 
precise evaluation of the integrals.

\subsection{Diagrams with fermion triangles}\label{triangle}

For the computation of the diagrams containing fermion triangles
(\fig{diag}d) we have
basically adopted the numerical methods described 
in~\cite{Ferroglia:2003yj}.
The mass of the external electron lines is set to zero  
except for the cases with internal $Z\,\gamma$ and $\gamma\,\gamma$, where 
$m_e$ is used to regularize infrared and collinear divergences. 
In the cases $Z\,\gamma$ and $\gamma\,\gamma$ we have also kept all the 
internal fermion masses ($b,t,\tau$), 
of the third generation,
while in the cases $W\,W$ and $Z\,Z$ all fermions are treated as  
massless, except the top quark.

The resulting expression for the contribution to $\sin^2\theta^{(2)}_{\rm eff}$
in (\ref{s2w}) can be written as a sum of terms
from tensor integrals of the family $V^{231}$ 
(the families of two-loop self-energies and vertices are collected in 
the appendix).
In order to simplify as much as possible the tensorial structure, we 
perform a simple reduction of the type
\bqa
\frac{2\,\spro{q}{p}}{(q^2+m^2)\,[(q+p)^2+M^2]}=
&{}&
\frac{1}{q^2+m^2} \,-\, \frac{1}{(q+p)^2+M^2}
\,-\,\frac{p^2-m^2+M^2}{(q^2+m^2)\,[(q+p)^2+M^2]} , \nonumber
\eqa
which does not introduce any new denominator and 
therefore any spurious singularity.
At this point we have a sum of tensor integrals up to rank 3 belonging to 
four vertex families ($V^{121}$, $V^{131}$, $V^{221}$, $V^{231}$), 
together with three self-energy types ($S^{111}$, $S^{121}$, $S^{131}$), 
and some one-loop diagrams ($B$ and $C$ functions).

\begin{figure}[ht]
\begin{center}
{\small
\noindent
\begin{tabular}{ccc}
\begin{picture}(100,50)(0,0)
\Line(0,0)(30,0)
\Line(90,34.5)(30,0)
\Line(90,-34.5)(30,0)
\Line(70,-23)(70,23)
\Text(10,4)[cb]{$p$}
\Text(86,-44)[cb]{$p_1$}
\Text(86,38)[cb]{$p_2$}
\Text(50,-23)[cb]{$m_1$}
\Text(80,-4)[cb]{$m_2$}
\Text(50,17)[cb]{$m_3$}
\end{picture}
\end{tabular}
\vspace{1.8cm}
}
\caption[]{One-loop three-point function. The momenta are 
understood as incoming.}
\label{C}
\end{center}
\end{figure}
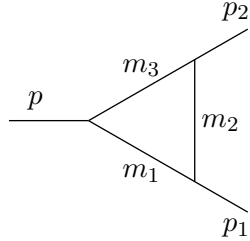

For all these tensor integrals, Feynman parameters are introduced.  
The parameter integrals after integration over the loop momenta  
are manipulated in order to obtain smooth 
integrands to be computed numerically. 
The methods used are those described in~\cite{Ferroglia:2003yj} together 
with some improvement aimed to increase the numerical stability in specific  
regions of the phase-space
(some new methods have been also developped for internal checks).
In this context,
it is worth mentioning the discovery of a new kind of algorithm for the 
computation of one-loop diagrams, which has been presented 
in~\cite{Uccirati:2004vy}.
The basic idea is that, given a polynomial $V$ in the Feynman parameters
$x=(x_1,x_2,\dots)^t$,
a constant $B$ and a column vector ${\cal P}$ with 
\bqa
V(x) &=& Q(x) + B,
\qquad\qquad {\cal P}^t\,\partial_x\,Q(x)= - Q(x),
\label{defQ}
\eqa
it can be easily proved that 
the following relation holds ($\beta > 0$),
\bqa
\int {\rm d}x \, V(x)^\mu &=& \int {\rm d}x \, 
\left( \beta  - {\cal P}^t\,\partial_x \right)\,
\int^1_0 {\rm d}y \,y^{\beta-1}\, \left[Q(x)\,y+B\right]^\mu \, .
\label{BTnew}
\eqa
Since any one-loop diagram can be expressed as an integral of a quadratic
polynomial 
to a negative power, the previous formula can be applied with $\mu<0$.
After an integration by parts, which cancels $\partial_x$, we can perform 
the integration in $y$. 
Then the procedure can be iterated with appropriate values of $\beta$ to 
obtain smooth integrals.
For example, for the scalar one-loop three-point function we obtain
($\varepsilon = 4-D$)
\bqa
C_0 &=&
\left( \frac{\mu^2}{\pi} \right)^{\varepsilon/2}\!\!
\egam{1+\frac{\varepsilon}{2}}\!
\intsxy{x_1}{x_2}\,V(x_1,x_2)^{-1-\varepsilon/2}.
\nl
&=&
\sum_{i=0}^{2}\,\frac{a_i}{2}\,\intsx{x_1}\,\frac{1}{V[i](x_1)-B}\,
\ln\frac{V[i](x_1)}{B}
+ {\cal O}(\varepsilon),
\label{C0}
\eqa
where
\bqa
&{}&
V(x) = x^t\,H\,x + 2\,K^t\,x + L,
\qquad
B = L - K^t\,H^{-1}\,K,
\qquad
X= - H^{-1}\,K,
\nl
&{}&
V[0](x_1)= V(1,x_1),
\qquad
V[1](x_1)= V(x_1,x_1),
\qquad
V[2](x_1)= V(x_1,0),
\nl
&{}&
a_0= 1-X_1,
\qquad
a_1= X_1-X_2,
\qquad
a_2= X_2, \qquad X=(X_1,X_2)^t \, .
\label{BX}
\eqa
The matrix $H$, the column $K$, and the constant $L$ are related to the 
physical quantities in the vertex diagram [\fig{C}]
as follows ($j,l = 1,2$),
\bqa
&{}&
H_{lj}= - \spro{p_l}{p_j},
\qquad
K_j= \frac{1}{2}( k_j^2 - k_{j-1}^2 + m_{j+1}^2 - m_j^2 ),
\qquad
L= m_1^2 - i\epsilon,
\nl
&{}&
k_0= 0,
\qquad k_1 = p_1, \qquad k_2 = p_1 + p_2 \, .
\eqa
This representation  for $C_0$ (which can be easily generalised for tensor 
integrals) is used to compute those two-loop diagrams that can be 
expressed as an integral of a one-loop three-point function, as in
$V^{221}$ and $V^{231}$ (see~\cite{Ferroglia:2003yj} for explicit expressions).
In such cases the masses and momenta of the $C$ functions depend on 
Feynman parameters, which have to be integrated over. 
It is therefore crucial that the zeros of the denominator 
$\left(V[i](x_1)-B\right)$ in (\ref{C0}) 
are compensated by the logarithm.

Another important feature of the calculation is the infrared problem.
Although the diagrams of \fig{diag}d are IR finite, 
some  individual integrals arising from the reduction 
in the $Z\,\gamma$ case are divergent. 
A typical case is the scalar part of 
the following diagram, which has the structure

\begin{figure}[ht]
\noindent
\begin{equation}
\begin{picture}(100,50)(0,0)
\Photon(0,0)(30,0){1.3}{7}
\Line(90,34.5)(70,23)
\Photon(70,23)(50,11.5){1.3}{5}
\Line(50,11.5)(30,0)
\Line(90,-34.5)(70,-23)
\Photon(70,-23)(50,-11.5){1.3}{5}
\Line(50,-11.5)(30,0)
\Line(50,-11.5)(50,11.5)
\Line(70,-23)(70,23)
\Text(5,-14)[cb]{$Z$}
\Text(54,22)[cb]{$\gamma$}
\Text(54,-32)[cb]{$Z$}
\end{picture}
=
\qquad
\begin{picture}(100,50)(0,0)
\Photon(0,0)(30,0){1.3}{7}
\Photon(70,23)(50,11.5){1.3}{5}
\Line(50,11.5)(30,0)
\Photon(70,-23)(50,-11.5){1.3}{5}
\Line(50,-11.5)(30,0)
\Line(50,-11.5)(50,11.5)
\Text(5,-14)[cb]{$Z$}
\Text(54,22)[cb]{$\gamma$}
\Text(54,-32)[cb]{$Z$}
\end{picture}
\!\!\!\!\!\!\!\!\!\!\!\!\!\!\!
\times\,\,\,
\begin{picture}(100,50)(0,0)
\Photon(0,0)(30,0){1.3}{7}
\Line(90,34.5)(70,23)
\Photon(70,23)(30,0){1.3}{10}
\Line(90,-34.5)(70,-23)
\Photon(70,-23)(30,0){1.3}{10}
\Line(70,-23)(70,23)
\Text(5,-14)[cb]{$Z$}
\Text(45,16)[cb]{$\gamma$}
\Text(45,-26)[cb]{$Z$}
\end{picture}
\!\!\!
+
\quad
{\rm finite\,\,\,terms} .
\end{equation}
\vspace{1cm}
\end{figure}

\noindent
For all the IR divergent integrals in the decomposition
we have extracted the infrared one-loop 
functions and verified analytically that the sum of them cancels.
The remaining finite terms are then again expressed in terms of smooth 
integrals and computed numerically.
Details of this treatment of IR divergences 
will be given elsewhere~\cite{vir}.

The algebraic handling and the numerical evaluation has been done 
in parallel by two independent computations. 
For the numerical integration
the NAG library D01GDF \cite{naglib} was used  
in one case and the CUBA library \cite{Hahn:2004fe} in the other one.
According to the features of the integrator, the contribution to the 
final result are summed up either in a single integral or split
into a sum of several terms. 
In the second case the total error has been computed as the sum of the 
errors of the different integrals.
Thanks to the use of more than one method for each diagram type, it has 
been possible to achieve a good numerical precision and make a severe 
cross-check of the calculations.

\section{Results}\label{results}
The input parameters for the evaluation of the final result are put together
in Tab.~\ref{tab:parameters}. 
They are chosen in accordance with 
\cite{Awramik:2004ge} in order to make an immediate comparison possible.
$M_W$ and $M_Z$ are the experimental values of 
the $W$- and $Z$-boson masses~\cite{PDG}, 
which are the on-shell masses. 
They have to be converted to the values in the pole mass scheme~\cite{Freitas:2002ja},
labeled as $\overline{M}_W$ and $\overline{M}_Z$, 
which are used internally for the calculation.
These quantities are related via $M_{W,Z} = \overline{M}_{W,Z}+ \Gamma^2_{W,Z}/(2 \;M_{W,Z})$. 
For $\Gamma_Z$ the experimental value (Tab.~\ref{tab:parameters}) and for $\Gamma_W$ the
theoretical value has been used, {\it i.e.}  
$\Gamma_W = 
3 \;G_{\mu} M^3_W/\left(2 \sqrt{2}\pi\right) 
\left(1+ 2 \alpha_s\left(M_W^2\right)/\left(3 \pi\right)\right)$
with sufficient accuracy.

\begin{table}[!htb]
\begin{tabularx}{16.cm}{l c  l}
\hline\hline
parameter & \hspace*{10.cm} &value\\\hline
$M_W$ && $80.426$ GeV\\
$M_Z$ && $91.1876$ GeV\\
$\Gamma_Z$ && $2.4952$ GeV\\
$m_t$ &&  $178.0$ GeV \\
$\Delta\alpha\left(M^2_Z\right)$ && $0.05907$\\
$\alpha_s\left(M^2_Z\right)$ && $0.117$\\
$G_\mu$ &&  $1.16637 \times 10^{-5}$\\
$\overline{M}_W$ && $80.3986$ GeV\\
$\overline{M}_Z$ && $91.1535$ GeV\\\hline\hline
\end{tabularx}
\caption {\small Input parameters entering our computation. $M_W$ and $M_Z$ are the experimental 
values 
of the $W$- and $Z$-boson masses, whereas $\overline{M}_W$ and $\overline{M}_Z$ are the calculated 
quantities in the pole mass scheme. }
\label{tab:parameters}
\end{table}

The results are given for $\Delta \kappa$, eq.~(\ref{kappa}), and are 
listed in
Tab.~\ref{tab:results} and Tab.~\ref{tab:res} for various masses of the Higgs boson.
Tab.~\ref{tab:results} contains also the one-loop result for comparison and the 
corresponding results obtained in~\cite{Awramik:2004ge}. 
A finite $b$-quark mass has been kept in the one-loop result.
The errors on our two-loop-results (in brackets)
are due to the uncertainty from the numerical integration. 
Our results are in full agreement with those given 
in~\cite{Awramik:2004ge} (last column of Tab.~\ref{tab:results}).

\begin{figure}[!htb]
\begin{tabular}{l}
\includegraphics[width=.9\linewidth]{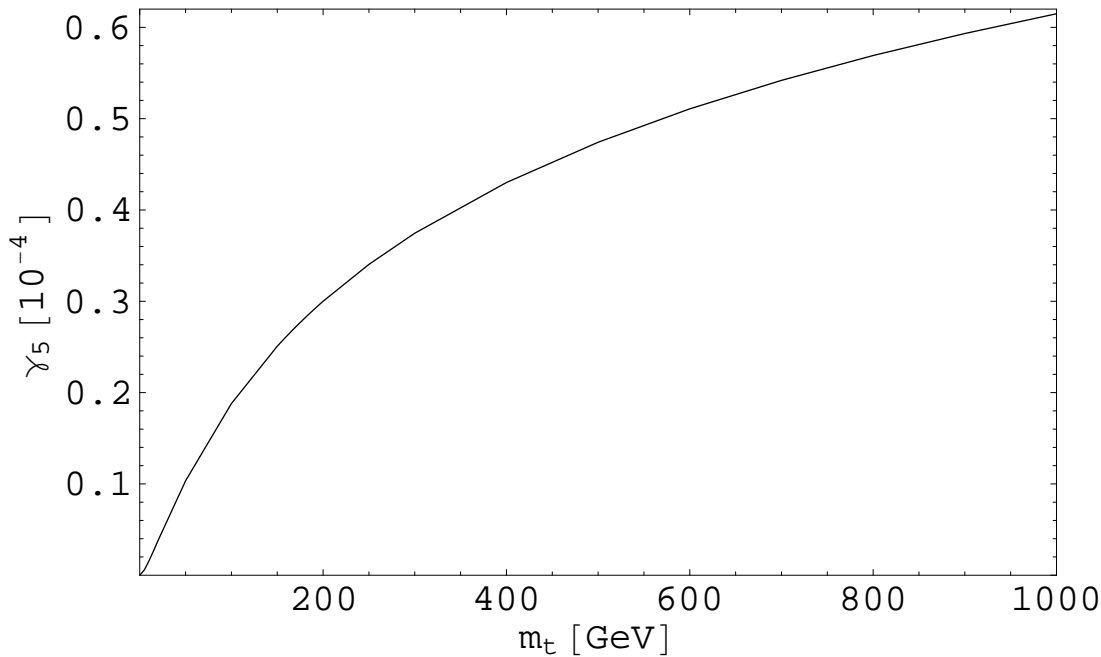}
\end{tabular}
\caption{"$\gamma_5$" contribution at $P^2=0$ as a functions of $m_t$.}
\label{gamma5MT}
\end{figure}

In Tab.~\ref{tab:res} the various parts of the two-loop result are shown. Large cancellations 
between the part containing two fermion loops and the part containing only one fermion loop occur.
Moreover, one can see that, depending on the value of $M_H$, the first term of~(\ref{split}) 
(column 4) is about $15$-$20$ times larger than the second one (column 5). 
Hence, the complete result for the renormalized $Z\ell\ell$ vertex at $P^2=M_Z^2$ can be  well 
approximated by the much simpler expression at $P^2=0$. The "$\gamma_5$" part is also 
very well approximated by its value at $P^2=0$, which is $0.280 \times 10^{-4}$. 
The "$\gamma_5$" part does not contain 
any terms proportional to $m^4_t$ or $m^2_t$; for illustration
it is plotted in Fig.~\ref{gamma5MT}, 
at $P^2=0$ as a function of the top mass.

\begin{table}[!htb]
\begin{tabularx}{16.cm}{c c c c}
\hline\hline
$M_H\left[GeV\right]$ & $\mathcal{O}\left(\alpha\right) \times 10^{-4}$&$\mathcal{O}\left(\alpha^2\right)\times 10^{-4}$&$\mathcal{O}\left(\alpha^2\right)\times 10^{-4}$\cite{Awramik:2004ge}\\\hline
100 &  \hspace*{1.5cm}438.937  \hspace*{1.5cm}&  \hspace*{1.5cm}-0.637(1)\hspace*{1.5cm}  &  \hspace*{1.5cm}-0.63\hspace*{1.5cm}\\
200  &  \hspace*{1.5cm}419.599  \hspace*{1.5cm}&  \hspace*{1.5cm}-2.165(1)  \hspace*{1.5cm}&  \hspace*{1.5cm}-2.16\hspace*{1.5cm}\\
600  &  \hspace*{1.5cm}379.560  \hspace*{1.5cm}&  \hspace*{1.5cm}-5.012(1)  \hspace*{1.5cm}&  \hspace*{1.5cm}-5.01\hspace*{1.5cm}\\
1000 &  \hspace*{1.5cm}358.619  \hspace*{1.5cm}&  \hspace*{1.5cm}-4.737(1)  \hspace*{1.5cm}&  \hspace*{1.5cm}-4.73\hspace*{1.5cm}\\\hline\hline

\end{tabularx}
\caption {\small Two-loop result for $\Delta \kappa$ in comparison with the one-loop result and the 
result in~\cite{Awramik:2004ge}.}
\label{tab:results}
\end{table}

\begin{table}[!htb]
\begin{tabularx}{16.cm}{c c c c c c}
\hline\hline
$M_H$& 2 ferm. loops & red. & $\hat{\Gamma}\left(0\right)$ & $\Gamma\left(M_Z^2\right)-\Gamma\left(0\right)$ &$ \gamma_5$\\
$\left[GeV\right]$&$\times 10^{-4}$&$\times 10^{-4}$&$\times 10^{-4}$&$\times 10^{-4}$&$\times 10^{-4}$
\\\hline
100 & \hspace*{0.5cm} 13.758 \hspace*{0.5cm} &\hspace*{0.5 cm} -0.722 \hspace*{0.5 cm}& \hspace*{0.5 cm} -14.903 \hspace*{0.5 cm}& \hspace*{0.5 cm} 0.959(1) \hspace*{0.5 cm}& \hspace*{0.5 cm} 0.271 \hspace*{0.5 cm}\\
200  & \hspace*{0.5cm} 13.758  \hspace*{0.5cm} &\hspace*{0.5cm} -0.688 \hspace*{0.5cm} & \hspace*{0.5cm} -16.465 \hspace*{0.5cm} & \hspace*{0.5cm} 0.959(1) \hspace*{0.5cm} & \hspace*{0.5cm} 0.271 \hspace*{0.5cm} \\
600  & \hspace*{0.5cm} 13.758  \hspace*{0.5cm} &\hspace*{0.5cm} -0.501 \hspace*{0.5cm} & \hspace*{0.5cm} -19.499 \hspace*{0.5cm} & \hspace*{0.5cm} 0.959(1) \hspace*{0.5cm}  & \hspace*{0.5cm} 0.271 \hspace*{0.5cm} \\
1000 & \hspace*{0.5cm} 13.758  \hspace*{0.5cm} &\hspace*{0.5cm} -0.386 \hspace*{0.5cm} & \hspace*{0.5cm} -19.339 \hspace*{0.5cm} & \hspace*{0.5cm} 0.959(1) \hspace*{0.5cm}  & \hspace*{0.5cm} 0.271 \hspace*{0.5cm} \\\hline\hline

\end{tabularx}
\caption {\small Breakdown of the two-loop result. The column "2 ferm. loops" gives all 
contributions with two closed fermion loops, whereas the other columns only contain results 
with one closed fermion loop. "red." refers to the 
reducible contribution  of the diagrams depicted in 
Fig.~\ref{gendiag}a and Fig.~\ref{gendiag}b plus the product of one-loop contributions in
(\ref{expand}). $\hat{\Gamma}\left(0\right)$ is the first term of (\ref{split}) and  
$\Gamma\left(M_Z^2\right)-\Gamma\left(0\right)$ the second one. 
In the last column the "$\gamma_5$"-contributions are given (see text).}
\label{tab:res}
\end{table}

\clearpage


\noindent
In conclusion, 
we have evaluated the electroweak 2-loop corrections to $\sin^2{\theta_{\rm eff}}$ 
with at least one closed fermion loop. Methods to calculate the appearing two-loop vertex functions 
have been described. A discussion of the various individual parts of the 
two-loop result was given and agreement with~\cite{Awramik:2004ge} was found.

\vspace*{2cm}
\noindent
This work was partially supported by the European Community's Human Potential Programme 
under contract HPRN-CT--2000-149 ``Physics at Colliders''.
S.U. would like to thank R.~Bonciani, A.~Ferroglia and G.~Passarino for 
useful discussions.

\clearpage

\appendix
\section{Two-loop self-energy and vertex families}
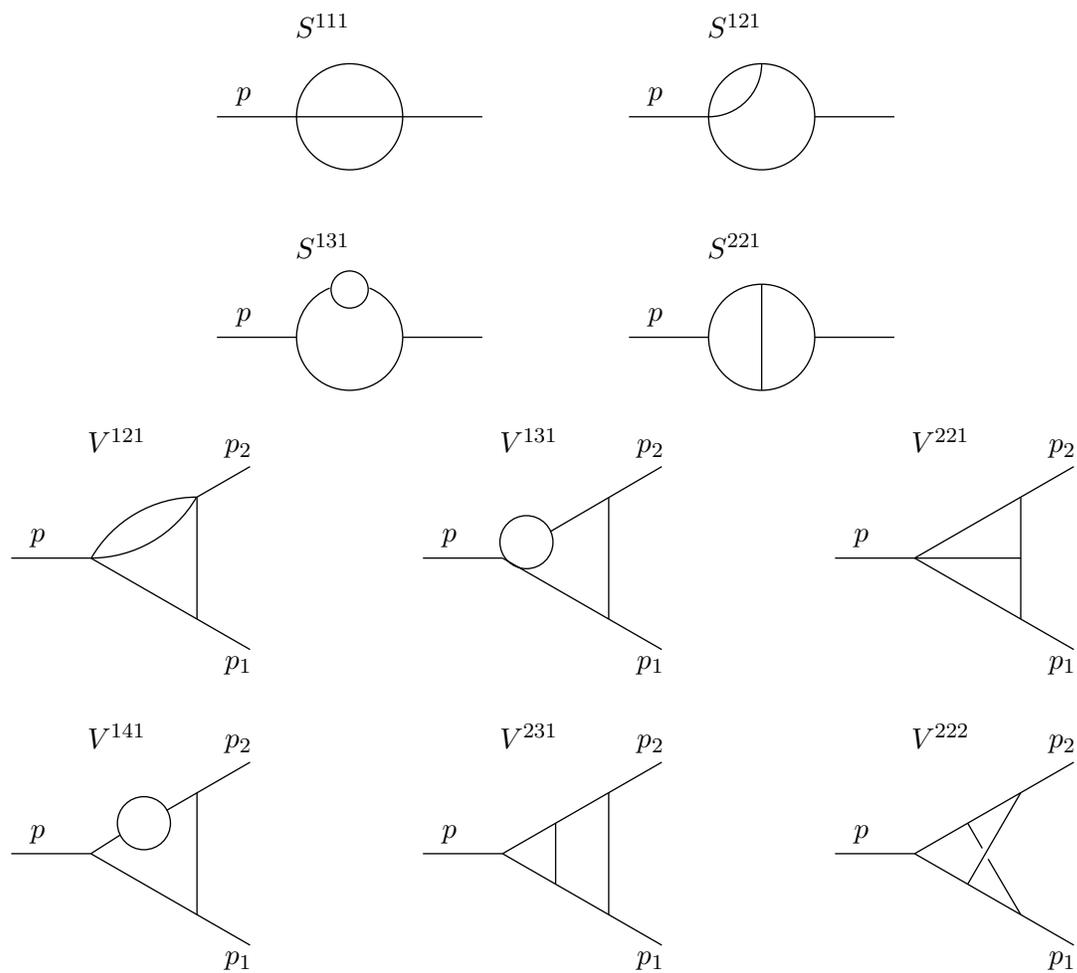
\begin{figure}[ht]
\begin{center}
{\small
\noindent
\begin{tabular}{cc}
\begin{picture}(100,50)(0,0)
\Line(0,0)(100,0)
\CArc(50,0)(20,0,360)
\Text(10,4)[cb]{$p$}
\Text(40,30)[cb]{$S^{111}$}
\end{picture}
&\qquad\qquad
\begin{picture}(100,50)(0,0)
\Line(0,0)(30,0)
\CArc(50,0)(20,0,360)
\CArc(30,20)(20,270,360)
\Line(70,0)(100,0)
\Text(10,4)[cb]{$p$}
\Text(40,30)[cb]{$S^{121}$}
\end{picture}
\end{tabular}
\vspace{1cm}

\begin{tabular}{cc}
\begin{picture}(100,50)(0,0)
\Line(0,0)(30,0)
\CArc(50,0)(20,0,68)
\CArc(50,0)(20,112,360)
\CArc(50,18)(7,0,360)
\Line(70,0)(100,0)
\Text(10,4)[cb]{$p$}
\Text(40,30)[cb]{$S^{131}$}
\end{picture}
&\qquad\qquad
\begin{picture}(100,50)(0,0)
\Line(0,0)(30,0)
\CArc(50,0)(20,0,360)
\Line(50,20)(50,-20)
\Line(70,0)(100,0)
\Text(10,4)[cb]{$p$}
\Text(40,30)[cb]{$S^{221}$}
\end{picture}
\end{tabular}
\vspace{1cm}

\begin{tabular}{ccc}
\begin{picture}(100,50)(0,0)
\Line(0,0)(30,0)
\Line(90,34.5)(70,23)
\CArc(70,-23)(46,90,150)
\CArc(30,46)(46,-90,-30)
\Line(90,-34.5)(30,0)
\Line(70,-23)(70,23)
\Text(10,4)[cb]{$p$}
\Text(86,-44)[cb]{$p_1$}
\Text(86,38)[cb]{$p_2$}
\Text(40,40)[cb]{$V^{121}$}
\end{picture}
&\qquad\qquad
\begin{picture}(100,50)(0,0)
\Line(0,0)(30,0)
\Line(90,34.5)(48,10)
\CArc(39,6)(10,0,360)
\Line(90,-34.5)(30,0)
\Line(70,-23)(70,23)
\Text(10,4)[cb]{$p$}
\Text(86,-44)[cb]{$p_1$}
\Text(86,38)[cb]{$p_2$}
\Text(40,40)[cb]{$V^{131}$}
\end{picture}
&\qquad\qquad
\begin{picture}(100,50)(0,0)
\Line(0,0)(70,0)
\Line(90,34.5)(30,0)
\Line(90,-34.5)(30,0)
\Line(70,-23)(70,23)
\Text(10,4)[cb]{$p$}
\Text(86,-44)[cb]{$p_1$}
\Text(86,38)[cb]{$p_2$}
\Text(40,40)[cb]{$V^{221}$}
\end{picture}
\end{tabular}
\vspace{2cm}

\noindent
\begin{tabular}{ccc}
\begin{picture}(100,50)(0,0)
\Line(0,0)(30,0)
\Line(90,34.5)(58.5,16.3)
\Line(41,7)(30,0)
\CArc(50,11.5)(10,0,360)
\Line(90,-34.5)(30,0)
\Line(70,-23)(70,23)
\Text(10,4)[cb]{$p$}
\Text(86,-44)[cb]{$p_1$}
\Text(86,38)[cb]{$p_2$}
\Text(40,40)[cb]{$V^{141}$}
\end{picture}
&\qquad\qquad
\begin{picture}(100,50)(0,0)
\Line(0,0)(30,0)
\Line(90,34.5)(30,0)
\Line(90,-34.5)(30,0)
\Line(50,-11.5)(50,11.5)
\Line(70,-23)(70,23)
\Text(10,4)[cb]{$p$}
\Text(86,-44)[cb]{$p_1$}
\Text(86,38)[cb]{$p_2$}
\Text(40,40)[cb]{$V^{231}$}
\end{picture}
&\qquad\qquad
\begin{picture}(100,50)(0,0)
\Line(0,0)(30,0)
\Line(90,34.5)(30,0)
\Line(90,-34.5)(30,0)
\Line(50,-11.5)(70,23)
\Line(70,-23)(57.7,-2)
\Line(55.5,2)(50,11.5)
\Text(10,4)[cb]{$p$}
\Text(86,-44)[cb]{$p_1$}
\Text(86,38)[cb]{$p_2$}
\Text(40,40)[cb]{$V^{222}$}
\end{picture}
\end{tabular}
\vspace{1.8cm}
}
\caption[]{Two-loop self-energy and vertex families. The momenta are 
all incoming and $p= -P= -\,(p_1+p_2)$.}
\label{vertices}
\end{center}
\end{figure}

\end{document}